\documentclass[journal]{IEEEtran}
\usepackage{amsmath,amsfonts}
\usepackage{algorithmic}
\usepackage{algorithm}
\usepackage{array}
\usepackage{textcomp}
\usepackage{stfloats}
\usepackage{url}
\usepackage{verbatim}
\usepackage{graphicx}
\usepackage{subfigure}
\usepackage{cite}
\usepackage{multicol}
\usepackage{color}

\usepackage{multirow}

\begin{document}
\title{SAR Ship Target Recognition Via Multi-Scale Feature Attention and Adaptive-Weighed Classifier}
%
%
%

\author{Chenwei Wang,~\IEEEmembership{Student Member,~IEEE,}
        Jifang Pei,~\IEEEmembership{Member,~IEEE,}
        Siyi Luo, 
        Weibo Huo,~\IEEEmembership{Member,~IEEE,} \\
        Yulin Huang,~\IEEEmembership{Senior Member,~IEEE,}
        Yin Zhang,~\IEEEmembership{Member,~IEEE,}
        Jianyu Yang,~\IEEEmembership{Member,~IEEE}
\thanks{This work was supported by the National Natural Science Foundation of China under Grants 61901091 and 61901090. (\emph{Corresponding author: Jifang Pei.})

       The authors are with the Department of Electrical Engineering, University of Electronic Science and Technology of China, Chengdu 611731, China (e-mail: peijfstudy@126.com; dbw181101@163.com).}}

\maketitle

\begin{abstract}
Maritime surveillance is indispensable for civilian fields, including national maritime safeguarding, channel monitoring, and so on, in which synthetic aperture radar (SAR) ship target recognition is a crucial research field.
The core problem to realizing accurate SAR ship target recognition is the large inner-class variance and inter-class overlap of SAR ship features, which limits the recognition performance. Most existing methods plainly extract multi-scale features of the network and utilize equally each feature scale in the classification stage. However, the shallow multi-scale features are not discriminative enough, and each scale feature is not equally effective for recognition. These factors lead to the limitation of recognition performance.
Therefore, we proposed a SAR ship recognition method via multi-scale feature attention and adaptive-weighted classifier to enhance features in each scale, and adaptively choose the effective feature scale for accurate recognition.
We first construct an in-network feature pyramid to extract multi-scale features from SAR ship images. 
Then, the multi-scale feature attention can extract and enhance the principal components from the multi-scale features with more inner-class compactness and inter-class separability. 
Finally, the adaptive weighted classifier chooses the effective feature scales in the feature pyramid to achieve the final precise recognition. 
Through experiments and comparisons under OpenSARship data set, the proposed method is validated to achieve state-of-the-art performance for SAR ship recognition.
\end{abstract}


%
\IEEEpeerreviewmaketitle

\section{Introduction}

\IEEEPARstart{M}{aritime} surveillance is indispensable for both military and civilian fields, including maritime disaster surveillance, channel monitoring, national maritime safeguarding, and so on \cite{intro1}. 
Ship monitoring plays a crucial and basic component in the system of maritime surveillance. Despite there existence of some transponder-based ship monitoring systems, like automatic identification systems (AIS) and vessel traffic services (VTS), they are inevitably problematic in the face of some unexpected or uncooperative situations.
Synthetic aperture radar (SAR) can provide high-resolution, weather-independent, day and night images, and thus, in recent years, serves for ship monitoring which mainly depend on ship detection and recognition. Compared with the rich research on ship detection, SAR ship recognition still needs more attention from the community.

Several pioneering studies have been conducted based on manually extracted features in recent years to solve the problem of SAR ship recognition. However, due to the fact that manually extracted features are inflexible and do not always work, these methods do not generalize well.
With the rapid development of machine learning, there are also many deep learning-based researches on SAR ship recognition problems \cite{inrto19,wang2022semi,wang2020deep,li2023panoptic,wang2022sar}.

Although some of these methods achieve state-of-art performances in SAR ship recognition, the crucial problem in practice, i.e., large inner-class variance and inter-class overlap of ship features, have not been adequately considered, leading to failure of recognition \cite{wang2021multiview,wang2019parking,wang2021deep,wang2020multi,intro5}. 
Since SAR ship images are usually grayscale intensity maps, the difference in the appearance of different ships is not obvious like the optical images, and there is a large interval of variation in the inner-class appearance, while a large inter-class overlap exists \cite{2compared21,wang2022recognition,liang2023efficient,wang2023entropy,2compared18,ai2021sar,compared2}.
For example, general cargo ships range in size from 90 to 200 meters long and 15 to 33 meters wide, while bulk carriers range in size from 150 to 275 meters long and 23 to 38 meters wide. The inner-class variance and inter-class overlap of these ships make ship recognition hard.

One of the effective ways to solve this problem is to use the multi-scale features of the ship \cite{2compared22,wang2023sar,wang2022global,chen2020ship}. However, most of the existing approaches extract multi-scale features plainly and utilize each feature scale equally at the classification stage. These shallow multi-scale features are often not discriminative enough, and each scale feature is often not equally effective for recognition, leading to the limitation in recognition performance.

\begin{figure*}
\centering
\includegraphics[width=0.85\textwidth]{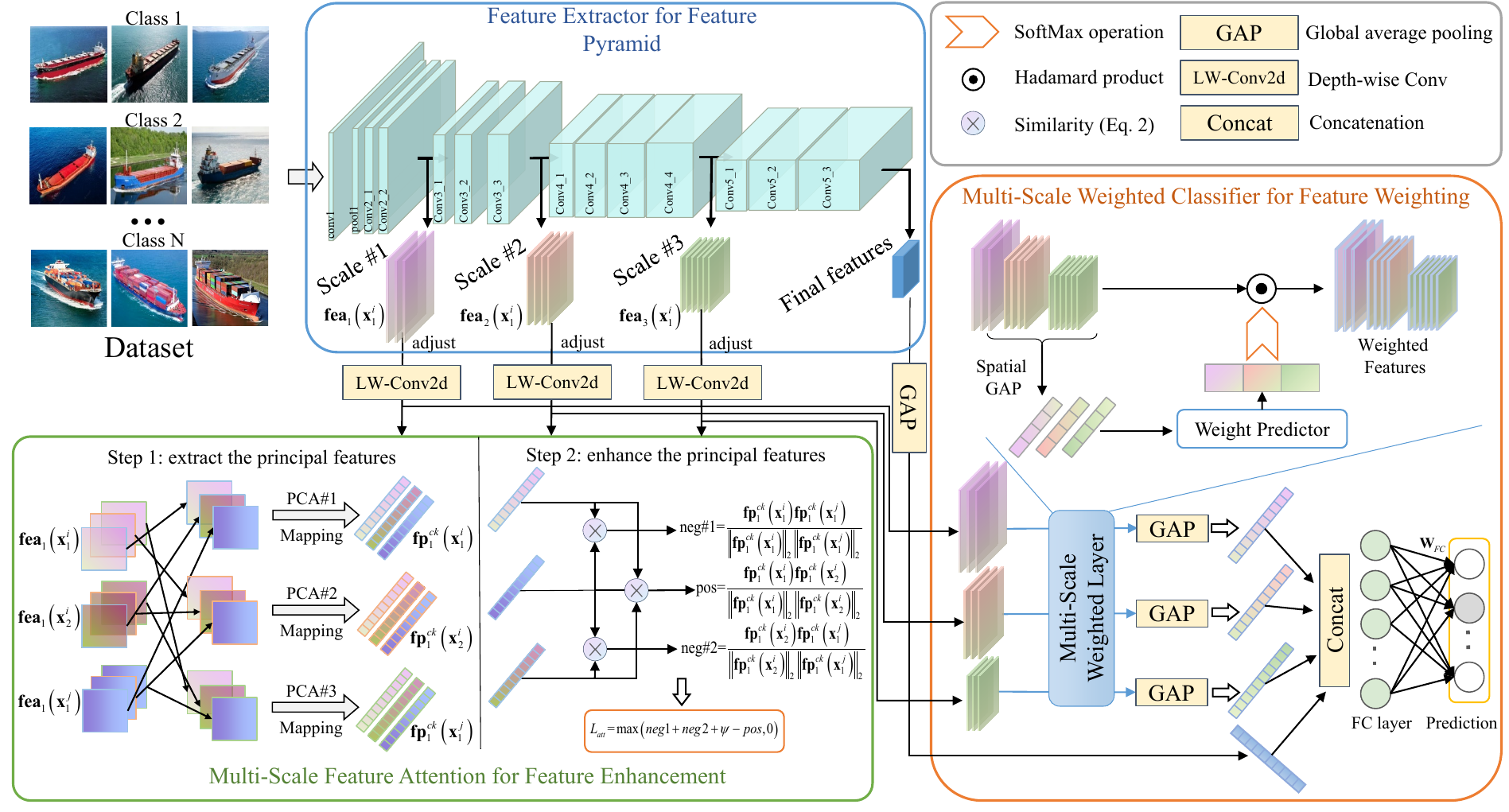}
\caption{Whole framework of proposed method.}
\label{framework}
\end{figure*}

Therefore, we propose a SAR ship recognition method via multi-scale feature attention and adaptive-weighted classifier, which optimally utilizes the multi-scale features to tackle the problem of large inner-class variance and inter-class overlap of SAR ship features. Our method first constructs an in-network feature pyramid to extract multi-scale features. Then the multi-scale feature attention enhances features with more inner-class compactness and inter-class separability on each scale. Finally, the adaptive-weighted classifier chooses and weights the effective feature scales, and discards other useless or ineffective feature scales to achieve accurate recognition. 

The main contributions of this letter are summarized as follows:

(1) One novel multi-scale feature attention is proposed to enhance the features with inner-class compactness and inter-class separability for better discrimination of SAR ship images. 

(2) One adaptive-weighted classifier is proposed to choose and weigh the effective feature scales. In this way, it can shift the major scales employed for accurate recognition.

(3) The proposed method achieves state-of-the-art recognition performances on OpenSARship data set with different numbers of training samples under 3-class and 6-class recognition experiments. 

The rest of this letter is organized below. Section \uppercase\expandafter{\romannumeral2} presents the framework and details of the proposed method. Section \uppercase\expandafter{\romannumeral3} verifies the effectiveness of the proposed method with experiments, and Section \uppercase\expandafter{\romannumeral4} gives our conclusion. 

\section{Proposed Method}
The proposed method consists of three main modules to address the problem of large inner-class variance and inter-class overlap of ships for recognition: 1) in-network feature pyramid to extract multi-scale features, 2) multi-scale feature attention for feature enhancement, 3) adaptive-weighted classifier for choosing effective scales. The details are introduced as follows. 

\subsection{Framework of Proposed Method}

As mentioned above,  the direct extraction and equal utilization of multi-scale features limited the recognition performance of SAR ship images. Therefore, our method enhances the inner-class compactness and inter-class separability of each feature scale, and then chooses and weighs the effective feature scales for the final recognition. As shown in Fig. \ref{framework}, the entire framework of the method consists of three main modules.

The first module, the feature extractor, constructs an in-network feature pyramid and provides several feature scales for the following enhancement and weighting of multi-scale features. 
The feature extractor is constructed by several residual blocks and gradually downsamples the scales of the feature maps to construct the feature pyramid. 
For each scale in the feature pyramid, the multi-scale feature attention enhances the principal features with inner-class compactness and inter-class separability to improve the discrimination of the shallow feature scales. 
The third module, the adaptive-weighted classifier, chooses the effective feature scales and improves the weights of the effective feature scales for the final precise recognition.
The pipeline of the framework can be described below. 

Given the input SAR ship images $\left\{ \mathbf{x}_{1}^{1},\mathbf{x}_{2}^{1},\ldots ,\mathbf{x}_{B}^{C} \right\}$, ${{\mathbf{x}}_j^i}$ is the $j$ SAR image in the $i$ class. The SAR ship images are inputted into the feature extractor to obtain several feature scales. Then these feature maps are fed into the multi-scale feature attention to improve their effectiveness and obtain the attention loss ${{L}_{att}}$. Finally, in the adaptive-weighted classifier, these features are weighted by a learnable weight predictor. After being concatenated with the final feature, they are inputted into the fully connected layer and Softmax to obtain the final recognition and recognition loss ${{L}_{recg}}$. The whole loss can be calculated as  
\begin{equation}
L={{\lambda }_{1}}{{L}_{att}}+{{\lambda }_{2}}{{L}_{recg}}
\end{equation}
where ${\lambda_1}$ and ${\lambda_2}$ are the weighting coefficients.

Through backpropagation and optimization, our method can be updated to achieve precise SAR ship recognition. The details of the multi-scale feature attention and adaptive-weighted classifier are as follows. 

\subsection{Multi-Scale Feature Attention}
To improve the feature discrimination in the shallow scales of pyramid network, the process of the multi-scale feature attention consists of two steps: 1) extracting the principal features, 2) enhancing these features by calculating the attention loss, as shown in Fig. \ref{framework}.

For simplicity, three scales of the feature are presented as an example. After going through the feature extractor and the light-weight convolutional layers to adjust the channel numbers of the feature maps, the features in three scales from ${{\mathbf{x}}_1^i}$ are set as $\mathbf{fea}\left( \mathbf{x}_{1}^{i} \right)\text{=}\left\{ \mathbf{fe}{{\mathbf{a}}_{1}}\left( \mathbf{x}_{1}^{i} \right),\mathbf{fe}{{\mathbf{a}}_{2}}\left( \mathbf{x}_{1}^{i} \right),\mathbf{fe}{{\mathbf{a}}_{3}}\left( \mathbf{x}_{1}^{i} \right) \right\}$. $\mathbf{fe}{{\mathbf{a}}_{1}}\left( \mathbf{x}_{1}^{i} \right)\in {{\mathbb{R}}^{{{h}_{1}}\times {{w}_{1}}\times {{c}_{1}}}}$ is the feature maps in the first scale of ${{\mathbf{x}}_1^i}$ and $h_1$, $w_1$ and $c_1$ means the numbers of the height, width, and channel. For simplicity, Fig. \ref{framework} shows the example of $c_1=3$ in the green block of multi-scale feature attention. 

At the first step, for each scale, the features of different SAR ship images from the same channel that are gathered together, like $\mathbf{fe}{{\mathbf{a}}_{1}}\left( \mathbf{x}_{1}^{i} \right)$ and $\mathbf{fe}{{\mathbf{a}}_{1}}\left( \mathbf{x}_{2}^{i} \right)$ are divided into $3$ groups of the feature maps. 
Then the principal component analysis (PCA) is employed to extract the principal feature vector of each group.

For the principal feature vector of different SAR ship images from the same channel, $\mathbf{fp}_{1}^{ck}\left( \mathbf{x}_{1}^{i} \right)$, $\mathbf{fp}_{1}^{ck}\left( \mathbf{x}_{2}^{i} \right)$, and $\mathbf{fp}_{1}^{ck}\left( \mathbf{x}_{1}^{j} \right)$, the similarity between every two vectors is calculated by
\begin{equation}
sim\left( \mathbf{x}_{1}^{i},\mathbf{x}_{2}^{i} \right)=\frac{\mathbf{fp}_{1}^{ck}\left( \mathbf{x}_{1}^{i} \right)}{{\left\| \mathbf{fp}_{1}^{ck}\left( \mathbf{x}_{1}^{i} \right) \right\|}_{2}} \cdot \frac{\mathbf{fp}_{1}^{ck}\left( \mathbf{x}_{2}^{i} \right)}{{\left\| \mathbf{fp}_{1}^{ck}\left( \mathbf{x}_{2}^{i} \right) \right\|}_{2}}
\end{equation}
where ${{\left\| \cdot  \right\|}_{2}}$ means the L2-norm. Accordingly, the similarity between the two vectors from different ship classes is denoted as $neg$, for instance, $neg1(i_1,j_1)=sim\left( \mathbf{x}_{1}^{i},\mathbf{x}_{1}^{j} \right)$ and $neg2(i_2,j_1)=sim\left( \mathbf{x}_{2}^{i},\mathbf{x}_{1}^{j} \right)$, and the similarity between the two vectors from same ship class is denoted as $pos$, for instance, $pos(i_1,i_2)=sim\left( \mathbf{x}_{1}^{i},\mathbf{x}_{2}^{i} \right)$. Finally, the attention loss ${{L}_{att}}$ can be calculated as 
\begin{equation}
{{L}_{att}}\text{=}\max \left( neg1+neg2+\psi -pos,0 \right)
\end{equation}
where $\psi $ means the margin of the vector similarity between the different classes and the same classes.

Through the process above, the multi-scale feature attention first extracts the principal feature vector in each scale. Then it enhances them with inner-class compactness and inter-class separability to improve the discrimination of SAR ship images in shallow scales of the feature pyramid.

\subsection{Adaptive-Weighted Classifier}
The adaptive-weighted classifier aims to choose the effective feature scales, and shift the major features employed for recognition among all the scales of the feature pyramid. The process of the adaptive-weighted classifier can be described as follows. 

Given the features in three scales from ${{\mathbf{x}}_1^i}$,
$\mathbf{fea}\left( \mathbf{x}_{1}^{i} \right)\text{=}\left\{ \mathbf{fe}{{\mathbf{a}}_{1}}\left( \mathbf{x}_{1}^{i} \right),\mathbf{fe}{{\mathbf{a}}_{2}}\left( \mathbf{x}_{1}^{i} \right),\mathbf{fe}{{\mathbf{a}}_{3}}\left( \mathbf{x}_{1}^{i} \right) \right\}$, the spatial GAP is employed to obtain the global feature vector $\mathbf{f}{{\mathbf{v}}_{1}}\left( \mathbf{x}_{1}^{i} \right)$, $\mathbf{f}{{\mathbf{v}}_{2}}\left( \mathbf{x}_{1}^{i} \right)$, and $\mathbf{f}{{\mathbf{v}}_{3}}\left( \mathbf{x}_{1}^{i} \right)$ from $\mathbf{fe}{{\mathbf{a}}_{1}}\left( \mathbf{x}_{1}^{i} \right)$, $\mathbf{fe}{{\mathbf{a}}_{2}}\left( \mathbf{x}_{1}^{i} \right)$, and $\mathbf{fe}{{\mathbf{a}}_{3}}\left( \mathbf{x}_{1}^{i} \right)$. The weight predictor is constructed by one fully connected layer and one batch-norm layer to obtain the corresponding weighting for all the features of different scales. 
The process above can be presented as 
\begin{equation}
\mathbf{f}{{\mathbf{v}}_{1}}\left( \mathbf{x}_{1}^{i} \right)=\text{SGAP}\left( \mathbf{fe}{{\mathbf{a}}_{1}}\left( \mathbf{x}_{1}^{i} \right) \right)
\end{equation}
\begin{equation}
weigh{{t}_{1}}\left( \mathbf{x}_{1}^{i} \right)=\text{WP}\left( \mathbf{f}{{\mathbf{v}}_{1}}\left( \mathbf{x}_{1}^{i} \right) \right)
\end{equation}
\begin{equation}
\mathbf{weight}=sf\left( \left[ weigh{{t}_{1}}\left( \mathbf{x}_{1}^{i} \right),
weigh{{t}_{2}}\left( \mathbf{x}_{1}^{i} \right),
weigh{{t}_{3}}\left( \mathbf{x}_{1}^{i} \right) \right] \right)
\end{equation}
\begin{equation}
\mathbf{fe{a}'}\left( \mathbf{x}_{1}^{i} \right)\text{=}\mathbf{fea}\left( \mathbf{x}_{1}^{i} \right)\odot \mathbf{weight}
\end{equation}
where $\text{SGAP}\left( \cdot  \right)$, $\text{WP}\left( \cdot  \right)$ and $sf\left( \cdot  \right)$ means the spatial GAP, the weighted predictor and the SoftMax, $ \odot$ means Hadamard product. 
Through the process above, the adaptive-weighted classifier finds out the effective feature scales for recognition and gives high weights to them for the following final precise recognition. 

The weighted features of different scales and the feature maps of the last layers in the feature extractor go through GAP and are concatenated together to obtain the final vector. By inputting the final vector into one fully connected layer and Softmax, the probability of the sample ${\mathbf{x}}_{1}^{i}$ classified to $j\text{th}$ class, ${{p}_{whole}}\left( {{y}_{j}}|\mathbf{x}_{1}^{i} \right)$, is given to calculate the recognition loss by
\begin{equation}
{{L}_{recg}}\left( \mathbf{x}_{1}^{i} \right)=-\sum\limits_{j=1}^{K}{{{y}_{j}}}\text{log}\left( {{p}_{whole}}\left( {{y}_{j}}|\mathbf{x}_{1}^{i} \right) \right)
\end{equation}

The proposed method first constructs a feature pyramid and provides multi-scale feature maps. Then, the multi-scale feature attention improves the discrimination of the shallow feature scales. Finally, the adaptive-weighted classifier optimally chooses the effective feature scales to achieve precise SAR ship recognition.
Furthermore, we conduct experiments to validate the effectiveness and practicability of the proposed method.

\section{Experimental Results}
In this section, the effectiveness of our method will be evaluated. We choose the measured benchmark dataset of SAR ship image, OpenSARShip, for the experiments. For the evaluation of practical application capabilities, the recognition experiments are run under decreasing labeled training SAR ship samples. 

\subsection{Dataset}
\begin{figure}[thb]
\centering
\includegraphics[width=1\linewidth]{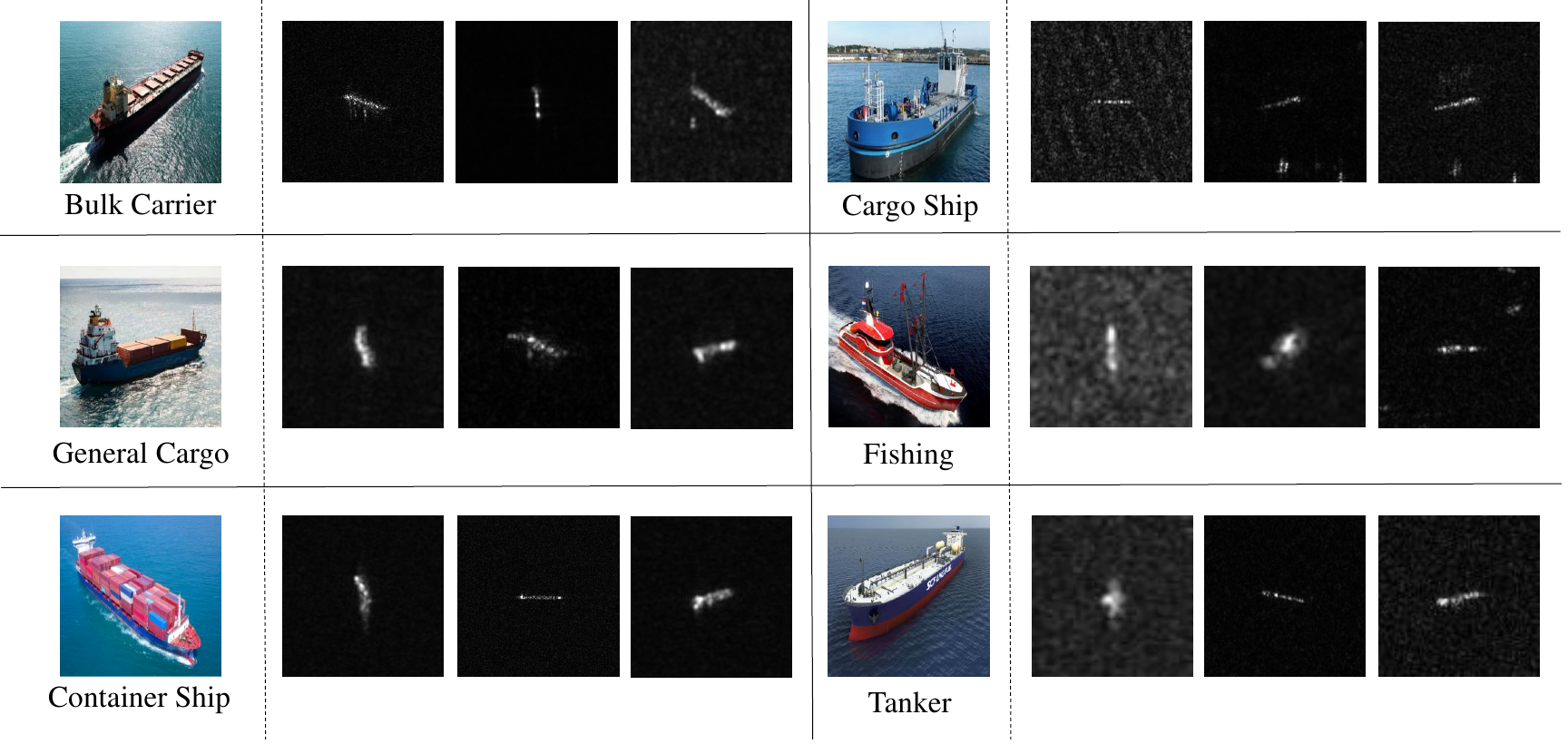} 
\caption{SAR ship images and corresponding optical ship images of six classes in OpenSARShip dataset.}
\label{sampleOPEN}
\end{figure}

The OpenSARShip dataset aims to develop complex ship detection and classification algorithms in a highly disturbed environment. These data were collected from 41 Sentinel-1 images under different environmental conditions. There are 11346 ship slices of 17 types of SAR ships and they are being integrated with AIS information. The labels of these ships are based on the AIS information, thus they are reliable \cite{ais}. Among our experiments, the ground range detected (GRD) data is used, which has a resolution of $2.0m \times 1.5m$ and a pixel size of $10m \times 10m$ in the azimuth and distance directions in Sentinel-1 IW mode. As for the dimensions of the ships, they range from 92m to 399m in length and from 6m to 65m in width. Six classes of SAR ship images are shown in Fig.\ref{sampleOPEN}.

\renewcommand{\arraystretch}{1.5}
\begin{table}[]
\centering
\caption{Image Number and Imaging Conditions of Different Targets in OpenSARShip}
\label{opensarset}
\setlength\tabcolsep{2pt}
\begin{tabular}{c|c|ccc}
\hline \hline 
Class          & Imaging Condition                                                                  & \begin{tabular}[c]{@{}c@{}}Training\\ Number\end{tabular} & \begin{tabular}[c]{@{}c@{}}Testing\\ Number\end{tabular} & \begin{tabular}[c]{@{}c@{}}Total\\ Number\end{tabular} \\ \hline 
Bulk Carrier   & \multirow{6}{*}{\begin{tabular}[c]{@{}c@{}} VH and VV, C ban\\  Resolution=$5-20$m\\ Incident angle=$20^{\circ}-45^{\circ}$ \\ Elevation sweep angle=$\pm 11^{\circ}$\\ ${\text{Rg20}}m \times {\text{az}}22m$\end{tabular}} & 200                                                       & 475                                                      & 675                                                    \\ \cline{1-1} \cline{3-5}
Container Ship &                                                                                    & 200                                                       & 811                                                      & 1011                                                   \\ \cline{1-1} \cline{3-5}
Tanker         &                                                                                    & 200                                                       & 354                                                      & 554                                                    \\ \cline{1-1} \cline{3-5}
Cargo          &                                                                                    & 200                                                       & 557                                                      & 757                                                    \\ \cline{1-1} \cline{3-5}
Fishing        &                                                                                    & 200                                                       & 121                                                      & 321                                                    \\ \cline{1-1} \cline{3-5}
General Cargo  &                                                                                    & 200                                                       & 165                                                      & 365     \\ \hline \hline                                               
\end{tabular}
\end{table}

\subsection{Recognition Performance and Comparison}
In this section, two types of recognition experiments are considered, 3 classes and 6 classes. Following \cite{open2,open5,2compared22}, we select three types of targets from OpenSARShip, i. e. , bulk carriers, container ships, and tanks. These three types of ships are the most common and typical ships, taking up 80\% of the international shipping market. Based on these three classes, we additionally select three more classes, namely cargo ships, fishing, and general cargoes, and organize a more challenging 6-class recognition experiment. The original datasets of the training and testing sets are shown in Table \ref{opensarset}. 

The numbers of targets in the six classes are unbalanced. To avoid the effect of class imbalance, we set the number of training samples for each class to be the same, and the specific value is determined by the training-testing ratio of 4:6 for the class with the smallest target sample size of the six classes. 
The number of different training numbers is augmented to 200 for each class by random sampling to reduce the influence of the number of different training data \cite{augment}.

\subsubsection{Recognition Performances of 3 Classes}
The SAR ship recognition performances of 3 classes in OpenSARShip under different training samples are presented in Table \ref{open3result}. It has illustrated that our method can achieve good performance when the training samples for each class are ranging from 100 to 20. 
Even when the training samples for each class are limited, like 30 or 20, our method still can achieve 70.37\% and 68.55\% respectively. These recognition performances of 3 classes in OpenSARShip illustrated that our method is effective and robust under a large range of the training sample for each class.

\subsubsection{Recognition Performances of 6 Classes}
The recognition ratios of 6 classes are shown in Table \ref{open6result}. The recognition of 6 classes is harder than that of 3 classes.
When the training samples for each class are 100 and 80, the recognition ratios achieve 59.16\% and 58.60\% respectively. If there are only 60 or 40 training samples for each class, our method achieves 57.35\% and 54.57\% respectively. Furthermore, when the training samples for each class continue to decrease to 30 or 20, the recognition of SAR ships becomes hard than in other conditions, our method still achieves 54.05\% and 52.75\%. 
The recognition performances of 3 and 6 classes have validated the effectiveness and robustness of our method. 
Furthermore, our method recognizes precisely containers and fishing, which is meaningful for combating illegal immigration and illegal fishing.

\renewcommand{\arraystretch}{1.5}
\begin{table}[]
\centering
\caption{Recognition Performance of 3 Classes with Different Number of Training Samples}
\label{open3result}
\resizebox{\linewidth}{!}{
\begin{tabular}{c|cccccc}
\hline \hline 
\multicolumn{1}{c|}{\multirow{2}{*}{Class}} & \multicolumn{6}{c}{Training Number in Each Class}          \\ \cline{2-7}
\multicolumn{1}{c|}{}                       & 20      & 30      & 40      & 60      & 80      & 100     \\ \hline 
Bulk Carrier                               & 53.16\% & 58.91\% & 69.25\% & 65.80\% & 65.23\% & 75.29\% \\
Container Ship                             & 73.70\% & 70.18\% & 68.64\% & 74.82\% & 79.75\% & 79.18\% \\
Tanker                                     & 75.20\% & 86.61\% & 86.61\% & 83.86\% & 87.01\% & 88.58\% \\ \hline 
Average                                    & 68.55\% & 70.37\% & 72.28\% & 74.18\% & 77.30\% & 79.97\% \\ \hline \hline 
\end{tabular} }
\end{table}

\renewcommand{\arraystretch}{1.5}
\begin{table}[]
\centering
\caption{Recognition Performance of 6 Classes with Different Number of Training Samples}
\label{open6result}
\resizebox{\linewidth}{!}{
\begin{tabular}{c|cccccccc}
\hline \hline
\multicolumn{1}{c|}{\multirow{2}{*}{Class}} & \multicolumn{6}{c}{Training Number in Each Class}          \\ \cline{2-7} 
\multicolumn{1}{c|}{}                       & 20      & 30      & 40      & 60      & 80      & 100     \\ \hline
Bulk Carrier                                & 54.95\% & 59.58\% & 53.05\% & 56.63\% & 65.68\% & 68.63\% \\
Container Ship                              & 73.61\% & 64.12\% & 68.56\% & 70.28\% & 70.16\% & 69.67\% \\
Tanker                                      & 40.96\% & 44.63\% & 48.59\% & 50.85\% & 53.39\% & 42.94\% \\
Cargo                                       & 29.62\% & 37.52\% & 39.50\% & 39.50\% & 40.93\% & 46.14\% \\
Fishing                                     & 69.42\% & 80.99\% & 80.99\% & 89.26\% & 83.47\% & 87.60\% \\
General Cargo                               & 34.55\% & 44.85\% & 34.55\% & 46.67\% & 33.94\% & 38.18\% \\ \hline 
Average                                     & 52.72\% & 54.05\% & 54.57\% & 57.35\% & 58.60\% & 59.16\% \\ \hline \hline
\end{tabular} }
\end{table}

\subsubsection{Comparison with Other Methods}
The comparisons with other methods are shown in Table \ref{3comparisonOPEN} and Table \ref{3&6COMPARISONOpenSARShip}.
Table \ref{3comparisonOPEN} is the comparison under different training SAR ship samples. 
Table \ref{3&6COMPARISONOpenSARShip} is the comparison with different effective deep learning networks under constant training SAR ship samples. CLSNet \cite{open2} and PFGFE-Net \cite{add1attention} in Table \ref{3&6COMPARISONOpenSARShip} is two representative methods with attention mechanisms. 

As shown in Table \ref{3comparisonOPEN}, it is clear that in the range of 100-20, our method can achieve the best recognition performances compared with other methods of SAR ship recognition. our method achieves 68.55\% under 20 samples for each class, Semi-Supervised \cite{open4} achieved 61.88\%, Supervised \cite{open4} achieved 58.24\%. When the training samples for each class are 80 or 100, the highest recognition ratios of other methods are 75.31\% under 100 and 68.67\% under 80, our method achieves 77.30\% under 80, and 79.97\% under 100. 

Table \ref{3&6COMPARISONOpenSARShip} shows the comparison of 3 classes and 6 classes with different deep learning networks.
The training samples of other methods of 3 classes and 6 classes are evaluated under 338 and 200 samples respectively for each class. The three indexes, Recall, Precision, and F1, can judge the model performance more comprehensively for a not-fully balanced dataset.
Under 3 classes, the accuracy of our method is only 0.72\% higher than the other best deep learning methods, the improvements in Recall, Precision, and F1 are more obvious, with 2.19\%, 3.28\%, and 2.36\% respectively.  
The same phenomenon can be observed from the recognition performance under 6 classes. These two performance comparisons have shown the superiority of our method.

From these recognition performances and comparisons above, it is clear that our method achieves the state-of-art performance of SAR ship recognition in a wild range of training samples for each class.

\renewcommand{\arraystretch}{1.3}
\begin{table}[]
\centering
\caption{Comparison of Performance (\%) of 3 classes with other methods of SAR ship recognition (The number in parentheses is the number of the training samples for each method)}
\label{3comparisonOPEN}
\begin{tabular}{c|cc}
\hline\hline
\multicolumn{1}{c|}{\multirow{2}{*}{Methods}} & \multicolumn{2}{c}{Number range of training images in each class}         \\ \cline{2-3} 
\multicolumn{1}{c|}{}                        & \multicolumn{1}{c|}{\makebox[0.27\linewidth]{1 to 50}}                                 & 51 to 338 \\ \hline
SRC\cite{SRC} &
  \multicolumn{1}{c|}{-} &
  71.81 (338)\\ \hline
LCKSVD\cite{LCKSVD} &
  \multicolumn{1}{c|}{-} &
  72.50 (338)\\ \hline
Semi-Supervised \cite{open4} &
  \multicolumn{1}{c|}{\begin{tabular}[c]{@{}c@{}}61.88 (20)\\ 64.73 (40)\end{tabular}} &
  \begin{tabular}[c]{@{}c@{}}68.67 (80) \\ 71.29 (120)\end{tabular} \\ \hline
Supervised \cite{open4} &
  \multicolumn{1}{c|}{\begin{tabular}[c]{@{}c@{}}58.24 (20)\\ 62.09 (40)\end{tabular}} &
  \begin{tabular}[c]{@{}c@{}}65.63 (80) \\ 68.75 (120)\end{tabular} \\ \hline
CNN\cite{compared2} &
  \multicolumn{1}{c|}{62.75 (50)} &
  68.52 (100)\\ \hline
CNN+Matrix\cite{compared2} &
  \multicolumn{1}{c|}{72.86 (50)} &
  75.31 (100) \\ \hline 
Proposed &
  \multicolumn{1}{c|}{\textbf{\begin{tabular}[c]{@{}c@{}} 68.55 (20) \\ 72.28 (40)\end{tabular}}} &
  \textbf{\begin{tabular}[c]{@{}c@{}} 74.18 (60) \\  79.97 (100)\end{tabular}} \\ \hline\hline
\end{tabular} 
\end{table}

\renewcommand{\arraystretch}{1.3}
\begin{table}[]\scriptsize 
\centering
\caption{comparison with effective deep learning networks under constant training samples. (The training samples of other methods under 3 classes and 6 classes are evaluated under 338 and 200 samples in each class. Ours is under 100 samples.)}
\label{3&6COMPARISONOpenSARShip}
\setlength{\tabcolsep}{1.15mm}
{
\begin{tabular}{c|cccc|cccc}
\hline \hline 
\multirow{2}{*}{Model} & \multicolumn{4}{c}{3 classes}                                                                                                                                                                                                                                                 & \multicolumn{4}{c}{6 classes}                          \\ \cline{2-9} 
                       & \begin{tabular}[c]{@{}c@{}}Recall \\ (\%)\end{tabular}                                                        & \begin{tabular}[c]{@{}c@{}}Precis- \\ ion(\%)\end{tabular}                                                    & \begin{tabular}[c]{@{}c@{}}F1 \\ (\%)\end{tabular}                                                            & \begin{tabular}[c]{@{}c@{}}Acc \\ (\%)\end{tabular}                                                           & \begin{tabular}[c]{@{}c@{}}Recall \\ (\%)\end{tabular}                                                        & \begin{tabular}[c]{@{}c@{}}Precis- \\ ion(\%)\end{tabular}                                                    & \begin{tabular}[c]{@{}c@{}}F1 \\ (\%)\end{tabular}                                                            & \begin{tabular}[c]{@{}c@{}}Acc \\ (\%)\end{tabular}   \\ \hline
ResNet-50               & 71.67                                                        & 66.79                                                        & 69.13                                                        & 72.82                                                        & 50.27  & 43.32     & 46.54 & 49.80 \\
DenseNet-169  & 71.40 & 68.83 & 70.07 & 74.31 & 55.55 & 47.21 & 51.07 & 54.26 \\
MobileNet-v3-L      & 65.12                                                        & 60.75                                                        & 62.84                                                        & 66.13                                                        & 49.95  & 42.14     & 45.71 & 46.60 \\
SqueezeNet-v1.1         & 67.42                                                        & 65.67                                                        & 66.45                                                        & 70.89                                                        & 52.72  & 43.73     & 47.81 & 50.83 \\
Inception-v4            & 69.26                                                        & 67.43                                                        & 68.28                                                       & 72.44                                                        & 54.92  & 46.46     & 50.34 & 54.55 \\
Hou et al. \cite{2compared17}            & 69.33                                                        & 69.44                                                        & 66.76                                                        & 67.41                                                        & 48.76  & 41.22     & 44.67 & 47.44 \\
Huang et al. \cite{2compared18}           & 74.74                                                        & 69.56                                                        & 72.04                                                        & 74.98                                                        & 54.09  & 47.58     & 50.63 & 54.78 \\
Xiong et al. \cite{2compared21}          & 73.87                                                        & 71.50                                                        & 72.67                                                        & 75.44                                                        & 53.57  & 45.74     & 49.35 & 54.93 \\
SF-LPN-DPFF \cite{2compared22}           & 78.83                                                        & 76.45                                                        & 77.62                                                        & 79.25                                                        & 54.49  & 48.61     & 51.38 & 56.66 \\ 
CLSNet \cite{open2}           & 77.87                                                        & 73.42                                                        & 75.05                                                       & 78.15                                                        & 54.20  & 46.66    & 50.15 & 53.77 \\
PFGFE-Net \cite{add1attention} & 76.03 & 77.80 & 76.91 & 79.84 & 53.93 & 44.61 & 48.83 & 56.83 \\
Proposed & \textbf{81.02} & \textbf{79.72} & \textbf{80.36} & \textbf{79.97} & \textbf{58.86} & \textbf{59.45} & \textbf{59.15} & \textbf{59.16}    \\ \hline \hline 
\end{tabular} }
\end{table}

\section{Conclusion}
In the field of SAR ship recognition, the problem of the large inner-class size variance and size inter-class overlap of SAR ships is hard to be solved. To tackle this problem, the existing methods proposed multi-scale features of SAR ships, but with the plain extraction and equal utilization for the recognition, limiting the recognition performance. 
Our method proposed a SAR ship recognition method via multi-scale feature attention and adaptive-weighted classifier to tackle this problem. 
The proposed method constructs an in-network multi-scale feature pyramid. The principal features are extracted and enhanced with inner-class compactness and inter-class separability. The effective feature scales for recognition are adaptively chosen and the weightings of these feature scales are improved for the final precise recognition.
The experimental results and comparisons under the OpenSARship dataset show that our method greatly improves the recognition performance under decreasing labeled training samples of ships which validates the practical application capabilities of our method.

\bibliographystyle{IEEEtran}
\bibliography{references}


%






\end{document}